# Thermal expansion in photo-assisted tunneling: visible light versus free-space terahertz pulses

Hüseyin Azazoglu, Rolf Möller, and Manuel Gruber[a]
*Faculty of Physics and CENIDE, University of Duisburg-Essen, 47057 Duisburg, Germany*

(Dated: 23 November 2023)

Photo-assisted tunneling in scanning tunneling microscopy has attracted considerable interest to combine sub-picosecond and sub-nanometer resolutions. The illumination of a junction with visible or infrared light, however, induces thermal expansion of the tip and the sample, which strongly affects the measurements. Employing free-space THz pulses instead of visible light has been proposed to solve these thermal issues while providing photo-induced currents of similar magnitude. Here we compared the impact of illuminating the same tunneling junction, reaching comparable photo-induced current, with red light and with THz radiations. Our data provide a clear and direct evidence of thermal expansion with red light-illumination, while such thermal effects are negligible with THz radiations.

Photo-assisted tunneling in scanning tunneling microscopy (STM) experiments allows investigating ultrafast dynamics on the scale of individual atoms and molecules. The focusing of optical pump and probe pulses at the STM junction leads to fast transients of the tunneling current, which allows stroboscopic measurements with sub-picosecond time resolution. However, the absorption of visible or infrared light at the STM junction causes local heating, which because of thermal expansion of the tip leads to a change of the tunneling gap[1–8]. As the photo-induced current is typically a small fraction of the total current, the optical beam is chopped at low-frequency, and the corresponding modulation of the current extracted with a lock-in. The lock-in signal is then generally dominated by the expansion and contraction of the tip rather than by the dynamics of the investigated system. This makes such measurements challenging, although particular implementations allow minimizing the impact of this issue[9–12].

Cocker and coworkers reported sub-picosecond and sub-nanometer resolutions by coupling free-space terahertz (THz) pulses to the STM junction[13]. The metal tip acts as an antenna focusing the electric field of the THz pulse to the STM junction, leading to fast voltage (and current) transients across the junction. The technique has recently attracted considerable attention and has been used to probe various dynamics[14–28]. In comparison to visible or infrared light, the THz fluence required to induce a tunneling current is smaller by orders of magnitude. Consequently, the heating and associated thermal expansion of the STM junction is expected to be negligible. Evidence for the absence of thermal artifacts is so far rather indirect: the average retraction of the STM tip evolves non-linearly with the THz-pulse energy[13], and the experimental data may be precisely reproduced from models omitting thermal effects[29].

Here, we compare the modulation of the tunneling current via the coupling of red light and of free-space THz pulses in the STM junction. Owing to the 100 MHz repetition rate of our setup, we can directly follow the time evolution of the THz-induced (and red-light-induced) current at various stages of a chopping cycle. The nature of the illumination (red light or THz) leads to profound differences in the time evolutions of the current. In addition, the responses to variations of the illumination power as well as to the polarity of the sample voltage are very different. Our measurements provide a direct evidence that thermal effects, *i.e.*, photo-induced expansion/contraction of the STM junction, are negligible with THz pulses.

We used a home-made variable-temperature STM operated in ultrahigh vacuum at room temperature. THz pulses with a repetition rate of 100 MHz are focused to the STM junction as described in Ref. 30. The light of a red laser diode is focused to the junction via an optical lense placed in front of another window of the ultra-high vacuum chamber. With this arrangement, we can conveniently compare the impact of red light and THz pulses on the same STM junction. Note that the junction was either illuminated with red light or with THz pulses, but never simultaneously with both sources. These two sources are electronically chopped. For all the measurements, the current feedback loop remained active but with the cutoff frequency far below the chopping frequency. Thereby, the response of the feedback to the induced current modulation is minimized while drifts occurring over longer times are corrected for. The measurements were performed on a Ag(111) surface with a tungsten tip (with a shaft diameter of 0.38 mm and length of ≈ 3.7 mm).

Figure 1 shows time traces of the tunneling current while the STM junction is illuminated with red light (red) and with THz radiations (black). Illumination with red light leads to a fast increase of the current (≈ 5 pA within ≈ 0.6 ms) followed by a slower, relatively linear increase. After the illumination, the current decays, again with a fast and slow responses. The almost linear and slow increase and decrease of the current upon intermittent illumination is consistent with a thermal expansion and contraction of the tip, which modulates the tip-sample distance. The fast response may be due to photo-current or a faster thermal dynamics of the STM junction. In contrast, for THz radiations, the evolution of the tunneling current (black curve in Figure 1) essentially resemble that of the chopping signal. The tunneling current is approximately 4 % higher when the THz radiation is active. Note that a detailed analysis reveals that the change in the detected current occurs within approximately 200 $\mu$s and is limited by the bandwidth of our home-made current-voltage transimpedance amplifier.

---

[a]Electronic mail: manuel.gruber@uni-due.de

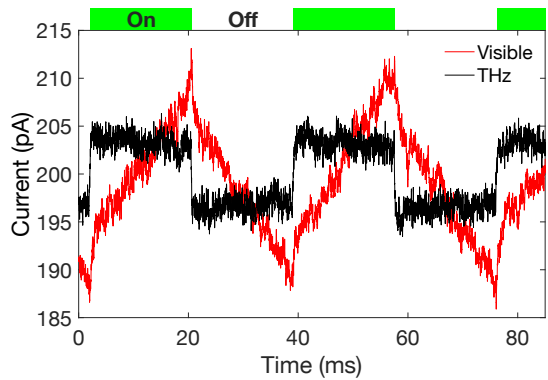

FIG. 1. Comparison of photo-assisted tunneling on a Ag(111) surface with visible red light (red) and with THz pulses (black) with a chopping frequency of 27 Hz. The green rectangles indicate the time during which illumination is active. The illumination of the STM junction with red light (nominal power of $\approx 5$ mW) leads to an almost linear increase of the current, which then decreases linearly in obscurity. In contrast, the junction illuminated with THz radiations essentially exhibits two current levels, reflecting the chopping of the THz emitters. The chopping frequency is 27 Hz, the current setpoint 200 pA. The sample voltage is 150 $\mu$V and 450 $\mu$V for the experiments with red and THz illuminations, respectively[31]. The displayed data were averaged over 155 traces.

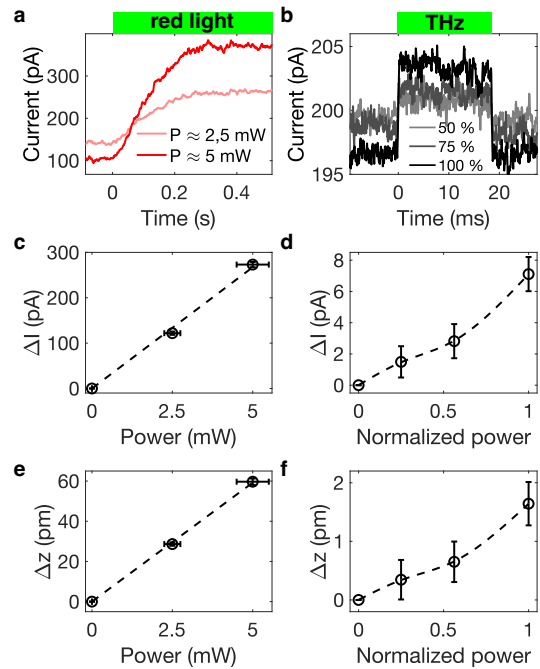

FIG. 2. Evolution of the tunneling current for various illumination powers. Time traces of the tunneling current upon illumination with **a** red light and **b** THz pulses. Evolution of the amplitude of the current modulation with **c** the red-light power and **d** the THz power. Variation of the tunnel barrier height as a function of the **e** red light and **f** THz power[32]. The THz power (in free space) is given relative to the maximal output power, corresponding to an estimated THz power of 40 $\mu$W. The chopping frequencies are **a** 0.5 Hz and **b** 27 Hz. The data are averaged over **a** 21 and **b** 155 sweeps.

The THz-induced modulation of the current has so far only been monitored via lock-in detection or by a change of the average tunneling current. The ability to follow the current within the subcycle of the THz chopping is here provided by the large repetition rate (100 MHz) of the laser combined with the usage of low DC sample voltages ($< 500\,\mu$V). As the THz-induced voltage across the junction is essentially independent of the tip-sample distance[29], smaller tip-sample distances (provided by the small DC voltage) lead to larger THz-induced currents.

Figure 2a shows the increase of the tunneling current as a function of time upon illumination with a red laser diode (nominal power of 5 mW). The current saturates after $\approx 300$ ms. With half of the nominal power, the response of the current has the same time constant, but the amplitude of the current modulation is approximately divided by two. The amplitude of the current modulation (Figure 2c) as well as the tip retraction (Figure 2e) scale linearly with the light power. This behavior is consistent with a thermally-induced linear evolution of the tip expansion with the illumination power[8]. The time-scale of thermalization of the tip is on the order of 100 ms.

The detected THz-induced current saturates in 200 $\mu$s (limited by the bandwidth of the transimpedance amplifier). As this is more than three order of magnitude faster than tip-thermalization, a thermal origin of the current modulation under THz illumination can safely be excluded. In addition, the evolutions of the current modulation amplitude and of the tip retraction are highly non-linear with the THz power (Figure 2b,d,f). This non linearity separately evidence a non-thermal origin of the photo-assisted tunneling. The evolution of the current amplitude with the THz power is essentially determined by the rectification of the current with THz pulses of various amplitudes, which primarily depends on the details of the $I$-$V$ characteristic of the junction[33].

The heating of the tip causes its elongation, and thereby a decrease of the tip-sample distance. This short distance should, for positive sample voltages, lead to an increased (positive) current, while more negative currents are expected for negative voltages. This behavior is observed for the junction illuminated with red light (Fig-ure 3a). The illumination increases the amplitude of the current. In contrast, the direction of the THz-induced current is given by (i) the direction of the electric field of the free-space THz pulses and (ii) the I-V characteristic of the junction over the voltage range spanned by the THz pulses. From a sample voltage of $+150$ to $-150\,\mu$V, none of these parameters are effectively changed[34], such that the direction of the THz-induced current should remain the same, as observed experimentally (Figure 3b). The sign of the DC current, however, is changed with the voltage polarity. The data of Figure 3 evidence, once more, the negligible thermal expansion of the tip upon illumination with THz radiations.

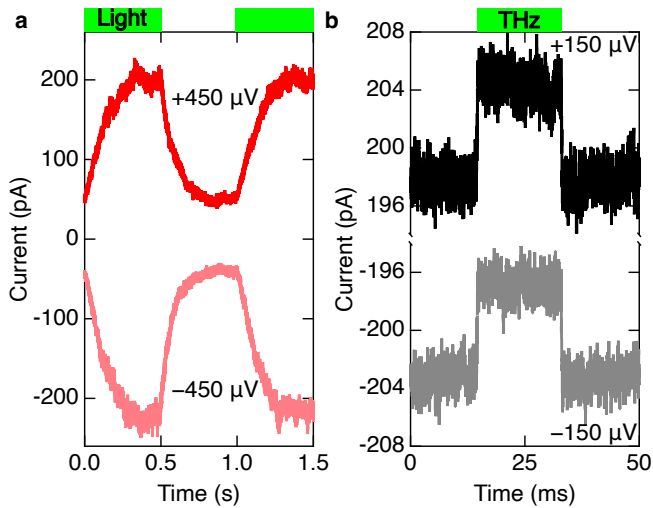

FIG. 3. Influence of the voltage polarity on the photo-assisted tunneling. Time traces of the tunneling current under illumination with **a** red light (5 mW) and **b** THz radiations for positive and negative DC voltages. The chopping frequencies are **a** 1 Hz and **b** 27 Hz. The data are averaged over **a** 12 and **b** 155 traces.

In conclusion, we compared the impact of red-light and THz illumination of a junction within the chopping cycle. Red-light induces heating and thereby an expansion of the tip, characterized by a time constant on the order of 100 ms, an amplitude proportional to the illumination power. This expansion induces an increase of the tunneling current amplitude. In contrast, the STM junction behaves completely differently with THz illumination. The time-constant of the chopped signal is on the order of $200\,\mu$s, limited by the bandwidth of our transimpedance amplifier. The amplitude of the THz-induced current is highly non-linear with the THz power, and the direction of the current is independent of the DC voltage polarity. We conclude that THz illumination allows photo-assisted tunneling with negligible thermal effects.

## ACKNOWLEDGMENTS

We thank Martin Mittendorff for fruitful discussions. Funding from the Deutsche Forschungsgemeinschaft (DFG; Project-ID 278162697 - CRC 1242, Project A08) is acknowledged.

## CONFLICT OF INTEREST

The authors have no conflicts to disclose.

## DATA AVAILABILITY

The data that support the findings of this study are available from the corresponding author upon reasonable request